\documentclass[twocolumn,aps,showpacs,prl]{revtex4}
\usepackage{amsmath}
\usepackage{graphicx}

\usepackage{dcolumn}
\usepackage{bm}
\usepackage{float}

\begin{document}

\title{Density-Temperature-Softness Scaling of the Dynamics of
Glass-forming Soft-sphere Liquids}
\author{ Pedro E. Ram\'irez-Gonz\'alez$^{1}$, Leticia L\'opez-Flores$^{2}$, Heriberto Acu\~na-Campa$^{3}$,
and Magdaleno Medina-Noyola$^{1}$}

\address{$^{1}$Instituto de F\'{\i}sica {\sl ``Manuel Sandoval Vallarta"},
Universidad Aut\'{o}noma de San Luis Potos\'{\i}, \'{A}lvaro
Obreg\'{o}n 64, 78000 San Luis Potos\'{\i}, SLP, M\'{e}xico \\ 
$^2$Facultad de Ciencias Fisico-Matem\'aticas, Benem\'{e}rita Universidad Aut\'{o}noma de Puebla, Apartado Postal J48,  72570, Puebla, M\'{e}xico.\\
$^{3}$ Departamento de F\'{\i}sica, Universidad de Sonora, Boulevard Luis
Encinas y Rosales, 83000, Hermosillo, Sonora, M\'{e}xico.}

\date{\today}

\begin{abstract}

We employ the principle of dynamic equivalence between soft-sphere and
hard-sphere fluids [Phys. Rev. E {\bf   68}, 011405 (2003)] to describe the interplay of the effects of varying the density $n$, the temperature $T$, and the softness (characterized by
a softness parameter $\nu^{-1}$) on the dynamics of glass-forming
soft-sphere liquids in terms of simple scaling rules. The main
prediction is the existence of a dynamic universality class associated with the hard sphere fluid, constituted by the soft-sphere systems whose dynamic parameters, such as
the $\alpha$-relaxation time and the long-time self-diffusion
coefficient,  depend on $n,\ T$, and $\nu$ only through the reduced
density $n^*\equiv n \sigma_{HS}^3(n,T,\nu)$, where the effective
hard-sphere diameter $\sigma_{HS}(n,T,\nu)$ is determined by the Andersen-Weeks-Chandler condition for soft-sphere--hard-sphere structural equivalence. A number of scaling
properties observed in recent experiments and simulations involving glass-forming
fluids with repulsive short range interactions are found to be a
direct manifestation of this general dynamic equivalence principle.
The self-consistent generalized Langevin equation (SCGLE) theory of
colloid dynamics is shown to accurately capture these scaling rules

\end{abstract}


\maketitle

The formation of colloidal glasses and gels has been the subject of
intense study during the last two decades \cite{sciortinotartaglia}. On the other hand, within their overwhelmingly rich phenomenology, molecular
glass-forming liquids exhibits intriguing universal features \cite{ngaireview1}. It has 
been natural to expect that the phenomenology of both, the glass
transition in ``thermally-driven" molecular glass formers, and the
dynamic arrest transition in ``density-driven" hard-sphere colloidal
systems, share a common underlying universal origin
\cite{jammingrheology}. In fact, interesting scalings of the
\emph{equilibrium} dynamics of simple models of soft-sphere glass formers, exposed by systematic computer simulations \cite{xu,berthierwitten1}, provide an initial clue to one possible physical origin of this universality, whose fundamental understanding, however, still constitutes an important theoretical challenge. This challenge  provides the main motivation of the present work.

An essential aspect of the equilibrium perturbation theory of liquids is the \emph{static structural equivalence} principle \cite{awc,verletweiss,hansen}. This principle states that a fluid at number concentration $n$ and temperature $T$, whose particles interact through a moderately soft and purely repulsive potential $u^{(\nu)}(r)$  (where $\nu^{-1}$ is some measure of softness), is structurally equivalent to a hard-sphere (HS) system with an effective HS diameter $\sigma_{HS}=\sigma_{HS}(n,T,\nu)$ and an effective volume fraction $\phi_{HS}=\phi_{HS}(n,T,\nu)= \pi n \sigma^3_{HS}(n,T,\nu)/ 6$. This means that the static structure factor (SSF) $S(k;n,T,\nu)$ of the soft-sphere system is given by $S(k;n,T,\nu)\approx S_{HS}(k\sigma_{HS};\phi_{HS})$, where $S_{HS}(k\sigma;\phi)$ is the SSF of the fluid of hard spheres of diameter $\sigma$ and volume fraction $\phi$. This static structural equivalence automatically implies the universality of the thermodynamic properties of the  class  of soft-sphere fluids defined, precisely, by this iso-structurality condition. 

The {\it dynamic} extension of this soft--hard equivalence was proposed more recently \cite{dynamicequivalence}, thus extending the referred thermodynamic universality to the dynamic domain, described by properties such as the \emph{self} intermediate scattering function (self-ISF) $F_S(k,t)$ or the long-time self-diffusion coefficient
$D_L$. Some implications of this universality, on the dynamic arrest scenario of soft-sphere systems, have also been discussed \cite{soft1}. In these discussions, however, temperature was considered constant, and hence, its role was never emphasized. 

The main purpose of the present work is to demonstrate that the same dynamic equivalence principle becomes a much deeper and more powerful fundamental tool when conceived as a general dynamic scaling principle in the density-temperature-softness state space of these glass-forming soft-sphere liquids. This defines what we refer to as the hard-sphere \emph{dynamic} universality class (HS-DUC) and explains, in particular, some of the intriguing scalings observed in the recent simulations on model glass-forming soft-sphere liquids \cite{xu,berthierwitten1}. 

Let us first refresh the concept of static structural
equivalence, now in terms of the radial distribution function (RDF)
$g(r;n,T,\nu)$ of a given soft-sphere model system. The fundamental physical notion is that this system behaves essentially as a hard-sphere system in the sense that $g(r;n,T,\nu) \approx
g_{HS}(r/\sigma_{HS};\phi_{HS})$
\cite{awc,verletweiss,hansen}, where $g_{HS}(r/\sigma;\phi)$ is the RDF of the HS system. This iso-structurality condition allows one to write
the equilibrium thermodynamic properties of the soft sphere system,
such as the equation of state $p=p(n,T,\nu)$, in terms of
the corresponding properties of the hard sphere fluid. For example,
the pressure $p(n,T,\nu)$  can be written, using the virial equation of
state \cite{hansen},  as $p(n,T,\nu)/nk_BT\equiv Z(n,T,\nu) \approx 1+4\phi_{HS}g_{HS}(1+;\phi_{HS})\equiv Z_{HS}(\phi_{HS})$. Using the Verlet-Weis prescription to approximate the contact value
$g_{HS}(1+;\phi_{HS})$ turns out to be equivalent to approximating
the hard sphere compressibility factor $Z_{HS}(\phi_{HS})$ by the Carnahan-Starling equation \cite{verletweiss}, thus
finally leading to the following approximate but universal mechanical equation of
state of the soft sphere system,
\begin{equation}
Z(n,T,\nu) \approx Z_{CS}(\phi_{HS})= \frac{1+\phi_{HS} +
\phi_{HS}^2 -\phi_{HS}^3}{(1-\phi_{HS})^3}, \label{virialoe}
\end{equation}
with $\phi_{HS}=\phi_{HS}(n,T,\nu)$ determined by the iso-structurality condition. This universality is nicely illustrated in the simulation results of Ref. \cite{xu} (see inset of Fig. 4 below).

The {\it dynamic} extension of this soft--hard equivalence was discussed in Refs. \cite{dynamicequivalence,soft1} in the context of the dynamics of  colloidal liquids, in which a short-time self-diffusion coefficient $D^0$ describes the diffusion of the colloidal particles ``between collisions". It is summarized by the statement that the self-ISF $F_S(k,t;n,T,\nu)$ of the fluid with soft repulsive potential $u^{(\nu)}(r)$ can be approximated by   \begin{equation}
\label{scalingforfself}  
F_S(k,t;n,T,\nu)\approx F^{(HS)}_S(k\sigma_{HS},D^0t/\sigma_{HS}^2;\phi_{HS}),
\end{equation}
where  $F^{(HS)}_S(k\sigma,D^0t/\sigma^2;\phi)$ is the self-ISF  of the fluid of hard spheres of diameter $\sigma$, volume fraction $\phi$, and (for simplicity) same short-time self-diffusion coefficient $D^0$ as the soft-sphere fluid. As a direct consequence of this dynamic universality, it follows that the long-time self-diffusion coefficient $D_L(n,T,\nu)\ (\equiv \lim _{t\to\infty}\langle (\Delta \textbf{r}(t))^2\rangle/6t)$ of the soft-sphere liquid, normalized as $D^*(n,T,\nu) \equiv D_L(n,T,\nu)/D^0$, is given by 
\begin{equation}
\label{scalingfordsl} D^{*}(n,T,\nu) \approx
D^{*}_{HS}[\phi_{HS}(n,T,\nu)],
\end{equation}
where $D^{*}_{HS}[\phi]$ is the corresponding property of the HS
system. Similarly, let us define the $\alpha$-relaxation time
$\tau_\alpha(k;n,T,\nu)$ by the condition
$F_S(k,\tau_{\alpha}) = 1/e$, which we normalize as
$\tau^*(k\sigma;\phi,T^*,\nu) \equiv
k^2D^0\tau_\alpha(k;n,T,\nu)$. The dynamic equivalence
principle above then implies that
\begin{equation}
\label{scalingfortau} \tau^{*}(k;n,T,\nu) \approx
\tau^{*}_{HS}[k\sigma_{HS}; \phi_{HS}],
\end{equation}
with $\tau^{*}_{HS}[k\sigma;\phi]$ refering to the HS system.

Some consequences of the universality summarized by Eq. (\ref{scalingforfself}) were illustrated in Refs. \cite{dynamicequivalence} and \cite{soft1} in the context of the truncated Lennard-Jones (TLJ) pair potential with tunable softness, $u^{(\nu)}(r)=\epsilon\left[(\sigma/r )^{2\nu}-2(\sigma/r)^\nu +1\right] \theta(\sigma-r)$ (with $\theta(x)$ being Heaviside's step function), whose state space is spanned by the volume fraction $\phi = \pi n \sigma^3/6$  and dimensionless temperature $T^*\equiv k_BT/\epsilon$. These references, however, discussed in detail only the limit of moderate softness ($\nu \gg 1$), in which the strong similarity with the HS potential leads to the additional simplification that $\sigma_{HS}(n,T,\nu)$ becomes $n$-independent, and given by the ``blip function" approximation \cite{hansen,soft1}. These, however, are actually unessential restrictions, and to  illustrate this we have performed Brownian dynamics simulations for $D_L$ of a  non-truncated and rather long-ranged soft repulsive potential (representative of highly charged colloids at low ionic strength), namely, the Yukawa potential $
u(r)/k_BT= K \exp [-z(r/\sigma-1)]/(r/\sigma)$ with $K=554$ and
$z=0.149$. 

\begin{figure}
\begin{center}
\includegraphics[scale=.2]{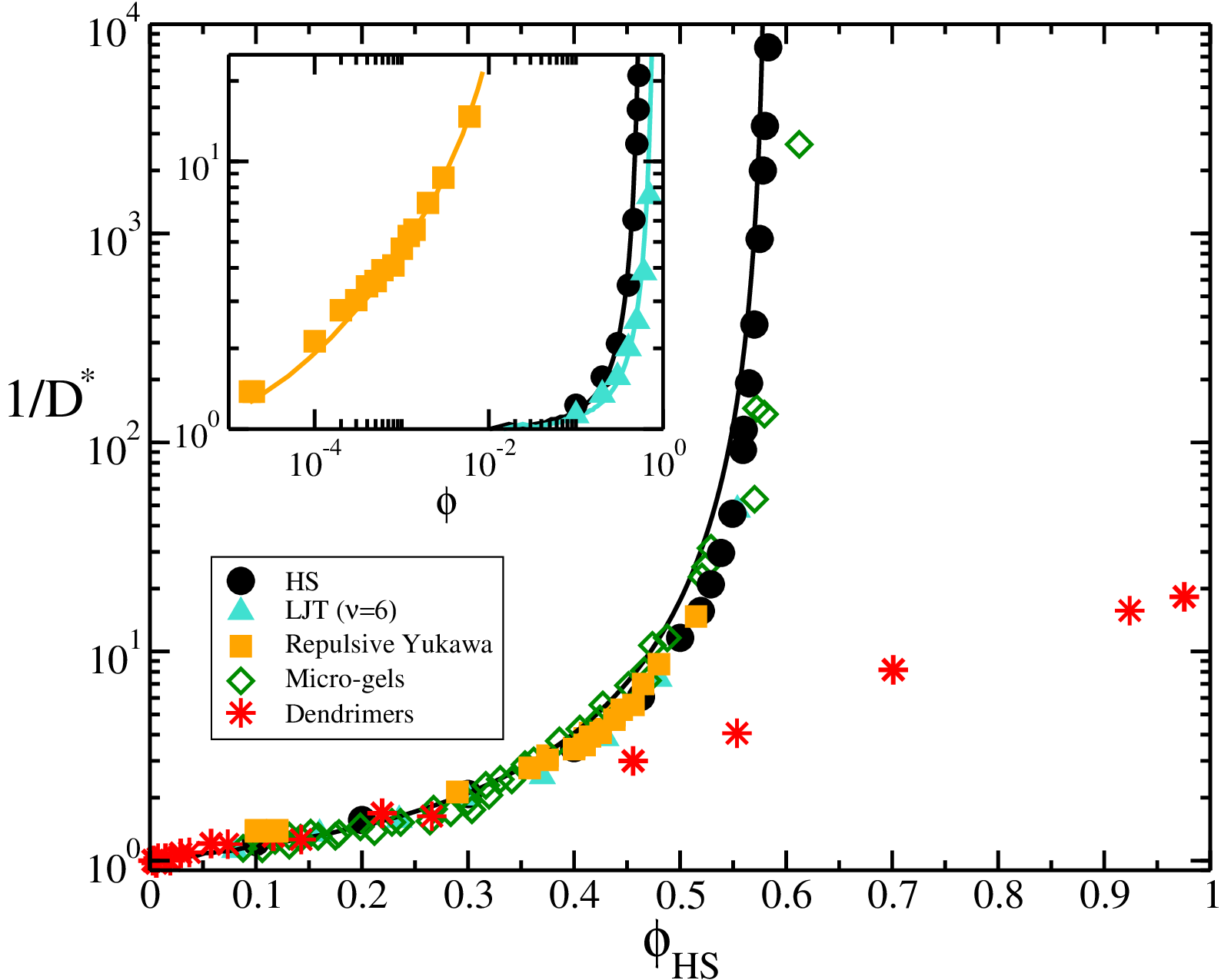}
\caption{Simulated data (solid symbols)  and SCGLE theoretical predictions (solid lines) of the normalized long-time self-diffusion coefficient $D^*(n,T,\nu)$, as a function of $\phi$ (inset) and as a function of $\phi_{HS} (n,T,\nu)$ (main figure) for the repulsive Yukawa fluid (squares), the truncated 6-12 Lennard-Jones fluid (triangles) and the hard sphere fluid (circles, Ref. \cite{gabriel}). The empty diamonds and the asterisks represent, respectively, the experimental master curve from Fig. 4 of Ref. \cite{trappe} for the reduced viscosity (proportional to $1/D^*$) of microgel solutions, and the data for $D_L(\phi)$ of third- and fourth-generation dendrimer solutions  in Fig. 3 of Ref. \cite{sagidullin}, scaled to collapse among themselves and with the HS master curve at low densities.} \label{fig1}
\end{center}
\end{figure}

These data are compared in Fig. 1 with the corresponding data for the TLJ system ($\nu=6$), much closer to the HS limit ($\nu=\infty$, Ref. \cite{gabriel}), also shown in the figure. For the three systems we plot $1/D^*$ as a function of   $\phi$ (inset) and of the effective HS volume fraction $\phi_{HS}(n,T,\nu)$, which is obtained not from with blip function method \cite{soft1}, but from the iso-structurality condition, written as $S(k_{max};n,T,\nu)=S_{HS}(k_{max}\sigma_{HS};\phi_{HS})$. This condition requests that the hight of the main peak of the static structure factor of the ``real" soft-sphere system, and of the effective hard-sphere system, coincide. As observed in the main figure, the data for $D^*$  of the three systems indeed collapse reasonably well when plotted as a function of $\phi_{HS}(n,T,\nu)$. Let us also notice that the self-consistent generalized Langevin equation (SCGLE) theory of colloid dynamics (Ecs. (1), (2), and (5-8) of Ref \cite{todos2},
with $k_c=1.35 k_{max}$), complemented with virtually exact liquid-theory approximations for $S(k)$, provides an excellent first-principles quantitative description of these data without any adjustable parameter (solid lines of Fig. \ref{fig1}).

In the same figure we have also included some experimental data for the relative low-shear viscosity  $\eta^*\equiv \eta(n)/\eta(n=0) $ of several microgel soft-sphere solutions reported in Ref. \cite{trappe}. These data correspond to samples with different soft-sphere size and softness which, upon a linear rescaling of the concentration, collapse onto a master curve (Fig. 2(a) of Ref.  \cite{trappe}). As illustrated in Fig. 1, such experimental master curve for $\eta^*(\phi)$ coincides pretty well with our HS simulation data for $1/D^*(\phi_{HS})$, thus indicating that these samples clearly belong to the hard-sphere dynamic universality class. One important question then refers to the conditions under which a system with arbitrary repulsive interaction will belong to this dynamic universality class. Our conjecture is that such condition is the existence of a distance of closest approach $\sigma_{min}$, such that $u^{(\nu)}(r)\gg k_BT$ for $r\le\sigma_{min}$, so that particle-particle overlaps are highly unlikely or forbidden. Thus, according to this conjecture, systems with ultrasoft repulsive interactions with finite overlap potential energy $u^{(\nu)}(0)$ will not belong to this HS dynamic universality class if $ k_BT\approx u^{(\nu)}(0)$. To illustrate this notion, in Fig. 1 we have also included the experimental data for $D^*$ of two low-generation dendrimer solutions, which cannot be adjusted by our HS universal curve, and whose structural properties are best described by the Gaussian core potential \cite{likos}.

\begin{figure}
\begin{center}
\includegraphics[scale=.2]{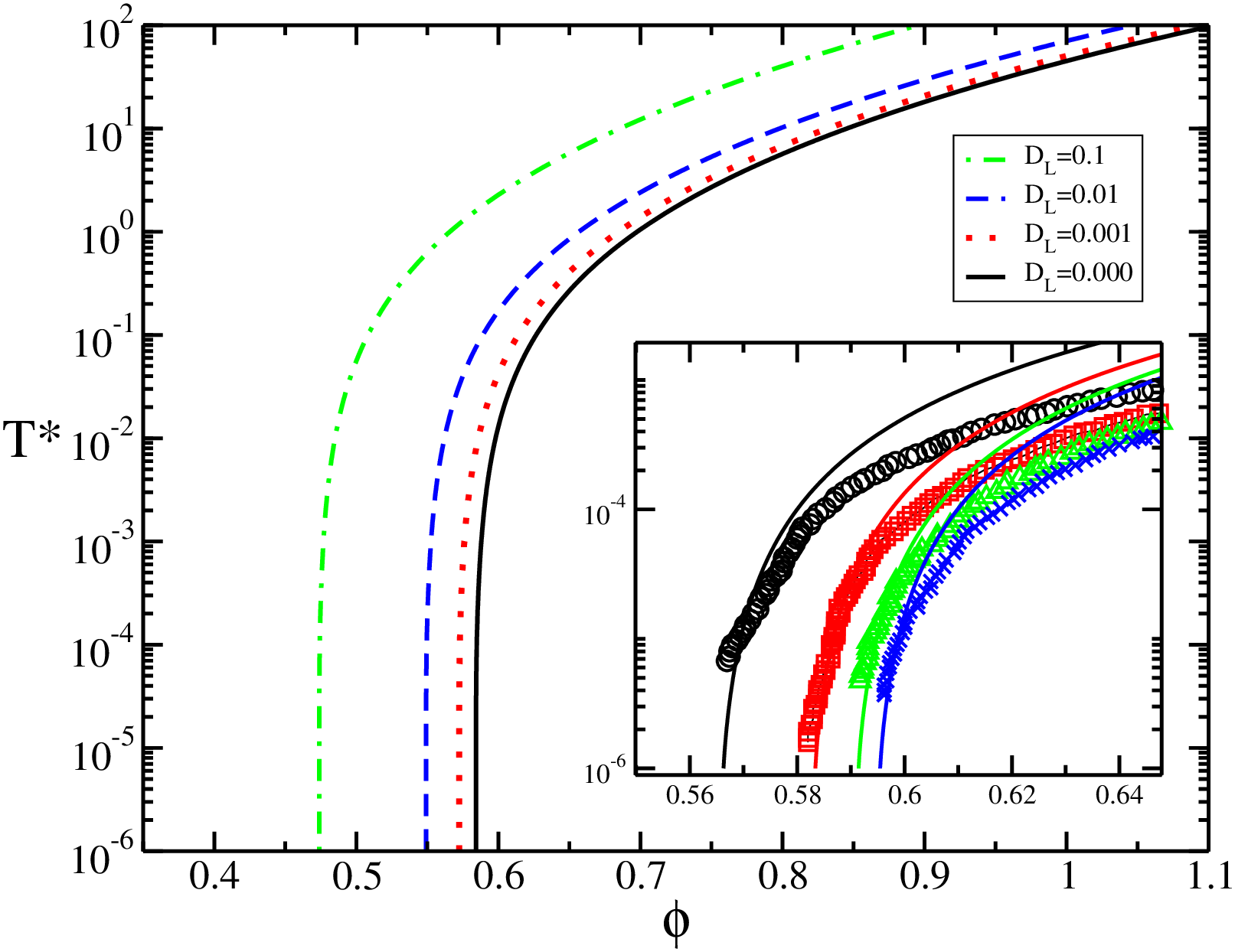}
\caption{State space of the truncated 6-12 Lennard-Jones fluid, showing four iso-diffusivity
lines, including the freezing transition (dash-dotted line) and
the ideal glass transition (solid line). In the inset we plot
similar iso-diffusivity lines for the soft-sphere system with
harmonic repulsive potential of Ref. \cite{berthierwitten1},
together with the simulation results for the iso-$\tau_\alpha$ lines labeled
$\tau_\alpha=10^1,10^2,10^3$,and $10^4$ in Fig. 1 of the same
reference. } \label{fig2}
\end{center}
\end{figure}

As illustrated in Fig. 1, the  SCGLE theory accurately captures the
scaling rules implied by Eqs. (\ref{scalingfordsl}) and
(\ref{scalingfortau}). In reality, most of the work leading to the results presented here was actually guided by the general qualitative predictions of this theory. For example,  from the general mathematical structure of the theory, it is clear that the dynamic equivalence principle just illustrated implies a far more general density-temperature-softness scaling, best
described in terms of the universal iso-dynamical surfaces in the
$(n,T,\nu)$ state space. These are defined as the loci of the
points with the same dynamical properties. According to Eqs.
(\ref{scalingfordsl}) and (\ref{scalingfortau}), the iso-diffusivity
surfaces labeled by the condition $D^*(n,T,\nu)=
\mathcal{D}^*$, with $\mathcal{D}^*$ being a prescribed constant
value, are also iso-$\tau^*$ and iso-$\phi_{HS}$ surfaces. Referring to the TLJ model system, in Fig.
\ref{fig2} we plot the cuts along the plane $\nu=6$ of these
iso-dynamical surfaces corresponding to $\mathcal{D}^*=10^{-1}, \
10^{-2}, \ 10^{-3}$ and 0.0 (calculated, however, as iso-$\phi_{HS}$
lines such that $\phi_{HS} (\phi,T^*,\nu)=0.494$, 0.55,
0.58, and 0.582, respectively). The correspondence between the labels
$\phi_{HS}$ and $\mathcal{D}^*$ was 
based, for the case of $\phi_{HS}=0.494$ (i.e., $\mathcal{D}^*\approx 10^{-1}$), on L\"owen's dynamic freezing criterion \cite{lowen}, and for the other iso-dynamical curves on the predictions of the SCGLE theory of colloid dynamics,
which locates the ideal glass transition at $\phi_{HS}= 0.582$
\cite{gabriel}.

Let us mention that it has recently been discovered \cite{atomicvscoloidal}
that the SCGLE theory of colloid dynamics
can also describe the long-time dynamics of atomic systems, provided
that the short-time self-diffusion coefficient $D^0$ assumes its
kinetic-theory value $D^0=(\sqrt{\pi}/16\phi)[\sigma\sqrt{k_BT/M}]$. This suggests the manner in which the
density-temperature-softness scaling and the iso-dynamical scenario
just discussed, may be shared by atomic systems. Notice, however,
that in this case $D^0$ becomes state-dependent, and hence, in
contrast with the Brownian case, we cannot expect that an
iso-$\tau^*$ surface will also be an iso-$\tau_\alpha$ surface. To
illustrate this, in the inset of Fig. \ref{fig2} we compare a few
SCGLE-predicted iso-$\tau^*$ lines for a soft-sphere fluid with
harmonic repulsive potential $u(r)=\epsilon (1-r/\sigma)^2$ for
$r\le\sigma$, with the iso-$\tau_\alpha$ lines determined by
\emph{molecular dynamics} simulations by Berthier and Witten
\cite{berthierwitten1} for this model system. In spite of the
expected quantitative differences, the theoretical and
simulated scenarios are, however, qualitatively identical.

Fig. \ref{fig1} provides the first illustration that the results for
$D^{*}_{HS}[\phi]$ actually play
the role of universal master curves  \emph{for all} moderately soft sphere
systems, provided that the horizontal axis is labeled not by the
volume fraction $\phi$ but by the \emph{effective} volume fraction
$\phi_{HS}(\phi,T^*,\nu)$. This scaling,
however, can be expressed in alternative manners. Thus, we can plot
$D^{*}_{HS}[\phi]$ (or $\tau^{*}_{HS}[k\sigma;\phi]$) not as a
function of $\phi$ but as a function of a combination of $\phi$,
such as $[\phi Z_{CS}(\phi)]$. The resulting curves will then also
be master curves for all the soft sphere systems in the HS dynamic universality class when the horizontal
axis is not the combination $[\phi Z_{CS}(\phi)]$ but the
combination $[\phi_{HS}(\phi,T^*,\nu)
Z_{CS}(\phi_{HS}(\phi,T^*,\nu))]$. 

\begin{figure}
\begin{center}
\includegraphics[scale=.2]{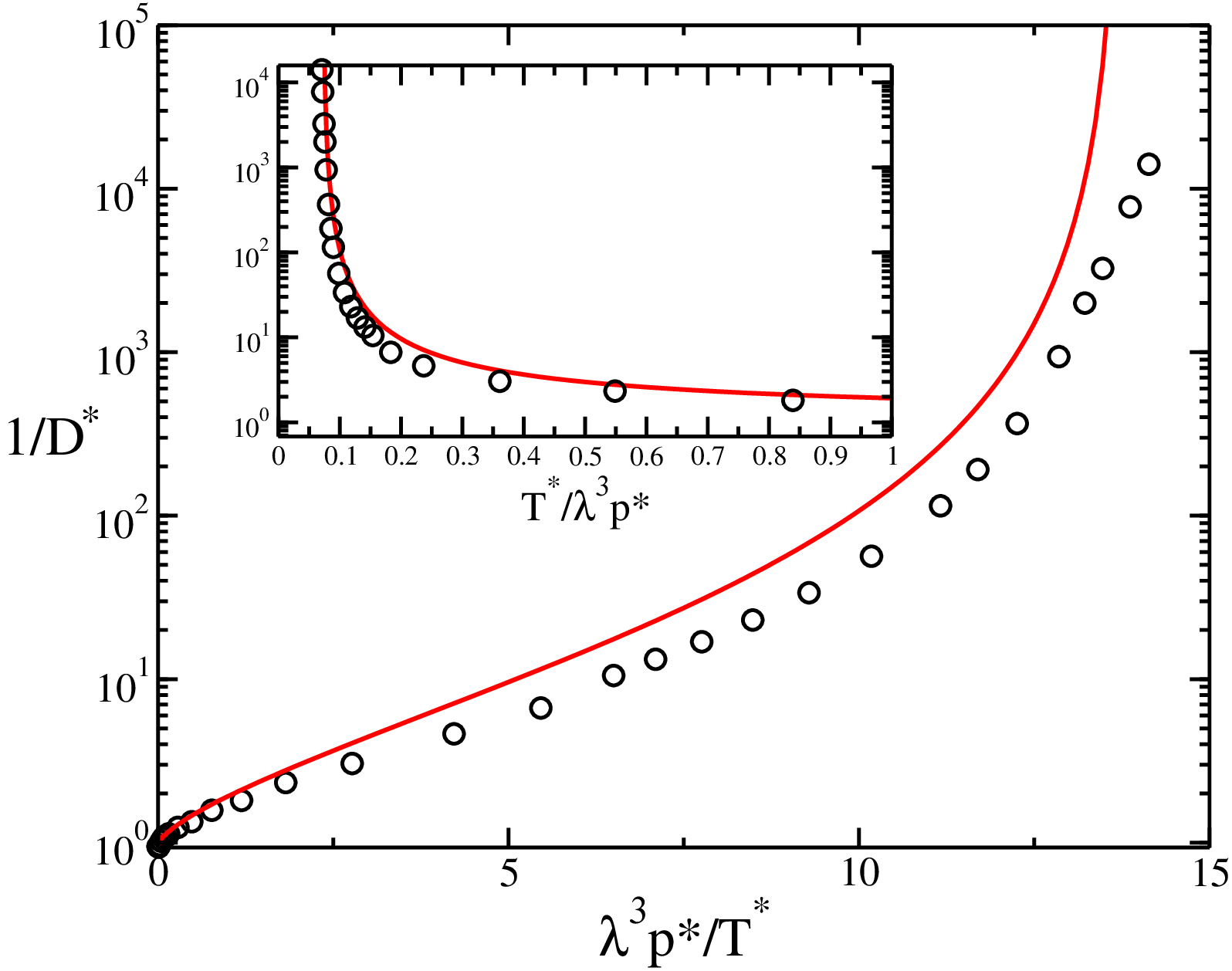}
\caption{Theoretical (solid line) and simulated (symbols) hard-sphere master curve for 
$D^*_{HS}(\phi_{HS})$ from Fig. \ref{fig1},
but now plotted as a function of the combination 
$[\phi_{HS}Z_{CS}(\phi_{HS})]=\lambda^3_\nu(T^*)p^*/T^*$ (main panel) 
and $T^*/(p^*\lambda^3_\nu(T^*))$ (inset). }
\label{fig3}
\end{center}
\end{figure}

A very
important meaning of the resulting master curves is suggested by the structure of the equation of state in Eq. (\ref{virialoe}), which can also be written in terms of
the dimensionless pressure $p^*_\nu \equiv
\pi\sigma^3p_\nu/6\epsilon$ as 
\begin{equation}\label{virialoereducedunits}
\frac{p^*_\nu(\phi,T^*)\lambda_\nu^{3}(\phi,T^*)}{T^*} =
\phi_{HS}(\phi,T^*,\nu) Z_{CS}(\phi_{HS}(\phi,T^*,\nu))
\end{equation}
with $\lambda_\nu(\phi,T^*)\equiv \sigma_{HS}(\phi,T^*,\nu)/\sigma$. Thus, the combination $[\phi_{HS} Z_{CS}(\phi_{HS})]$ is also the
combination $[p^*\lambda^3_\nu/T^*]$. In Fig. \ref{fig3} we
replot the master curve $D^{*}_{HS}[\phi]$ of Fig.
\ref{fig1} but now as a function of this combination. Alternatively,
we can use the combination $[\phi Z_{CS}(\phi)]^{-1}$, which changes
the horizontal axis to $[T^*/p^*\lambda^3_\nu]$. In the inset
of Fig. \ref{fig3} we replot the same results for $D^*_{HS}[\phi]$ as a function of this combination.
The resulting curves in Fig. \ref{fig3}  then
constitute a prediction of the existence of master curves that
describe the universal dependence of  $D^*(\phi,T^*,\nu)$ on pressure and temperature for this class of soft-sphere systems. The prediction of similar master curves for $\tau^*(k\sigma;\phi,T^*,\nu)$ can be drawn from the determination of the results for $\tau^*_{HS}(k\sigma;\phi)$.

\begin{figure}
\begin{center}
\includegraphics[scale=.2]{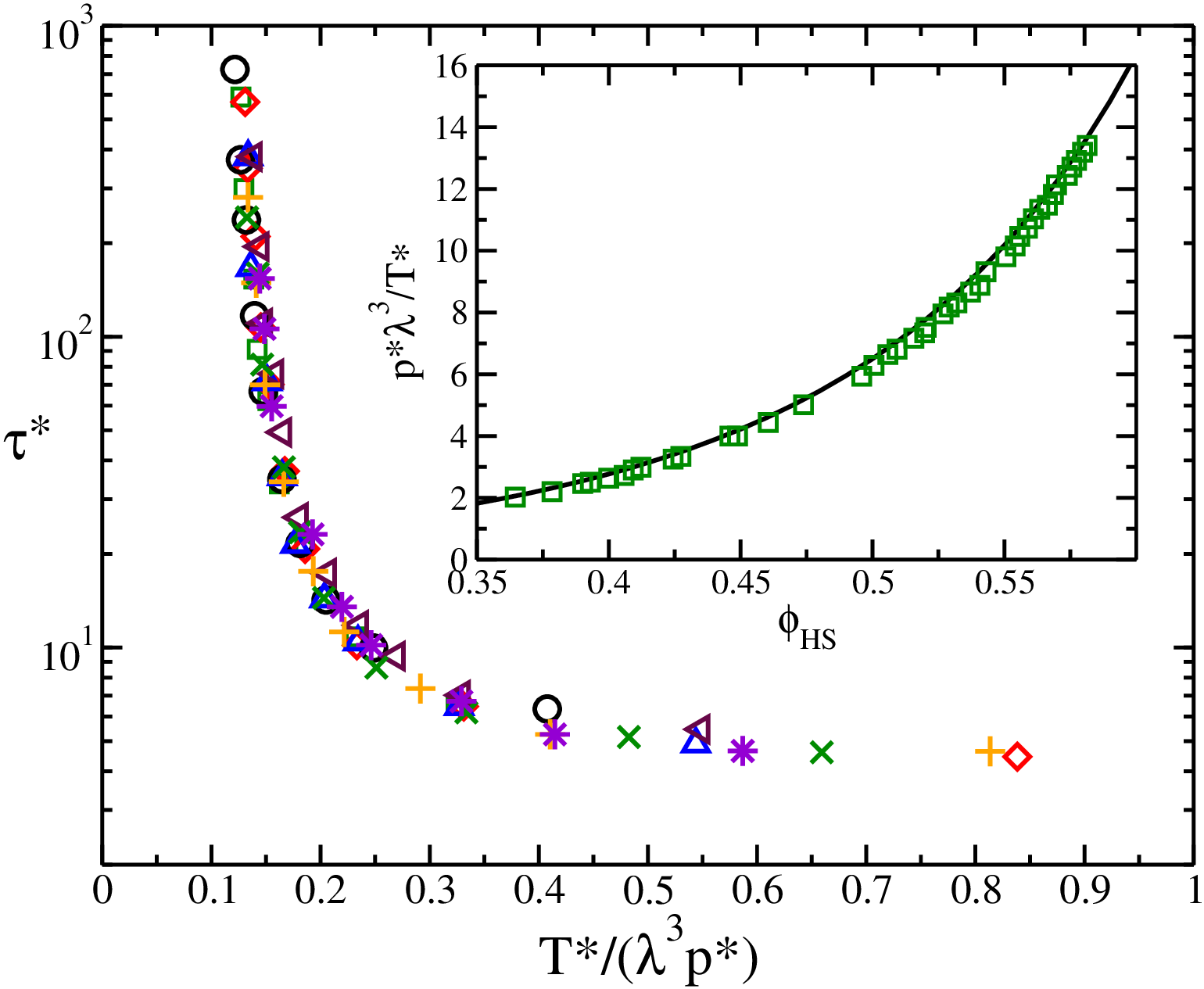}
\caption{Molecular dynamics 
simulations reported in Figs. 1a and 1b of Xu \emph{et\ al.} \cite{xu} for $\tau_\alpha$ of soft-sphere liquids, plotted here as the master curve for $\tau^* (T^*,p^*)$ as a function of the combination 
$[\phi_{HS}(\phi,T^*,\nu) Z_{CS}(\phi_{HS}(\phi,T^*,\nu))]^{-1}
=[T^*/\lambda^3_\nu(T^*)p^*]$.  In the inset we show the simulation data, obtained by 
Xu \emph{et.al.}, for the $p^*/T^*$ ratio (squares), compared with Eq. (\ref{virialoereducedunits}) 
(solid line).}
\label{fig4}
\end{center}
\end{figure}

This predicted scaling has actually been corroborated by the master curve
empirically discovered by Xu et al. \cite{xu} in their recent
simulations performed on the TLJ soft sphere system with $\nu=6$ and on other soft potentials. These authors found that the simulated results for the
$\alpha$-relaxation time $\tau_\alpha$ of these soft-sphere systems,
computed as a function of temperature $T$ at fixed pressure $p$, or
as a function of $1/p$ at fixed $T$ (Figs. 1.a and 1.b of Ref.
\cite{xu}, respectively) all collapse onto a master curve when
plotted as a function of the ratio $T/p$ (Figs. 2 and 3 of
\cite{xu}). The simplest manner to relate their empirical scaling
with our predictions in Fig. \ref{fig3} above, is to scale the raw
data for $\tau_\alpha$ in their Fig. 1 as $\tau^*
= k^2 D^0 \tau_\alpha$, with $D^0= (\sqrt{\pi}/16\phi)
[\sigma\sqrt{k_BT/M}]$, and to plot them as a function of
$[T^*/p^*\lambda^3_\nu(T^*)]$.  As a result, the various data of Fig.
1 of Xu et al. \cite{xu} collapse onto the master curve shown here
in Fig. \ref{fig4}, predicted by the dynamic equivalence
discussed in the present work, and in agreement with the scaling
discovered by Xu et al. This required to express $\phi$ in
this expression for $D^0$ in terms of $T^*$ and $p^*$, but this is
easily achieved using the equation of state in Eq.
(\ref{virialoereducedunits}); the inset of Fig. \ref{fig4} compares our
theoretical equation of state (Eq. (\ref{virialoereducedunits}))
with the corresponding simulation data of Ref. \cite{xu}. Let us
finally notice that for the TLJ model, in the low-temperature regime $\lambda_\nu (\phi,T^*)$ may be approximated by the blip-function result \cite{hansen,soft1}, which  yields $\lambda_\nu^3 (T^*)\approx
1-(3\sqrt{\pi}/2\nu)\sqrt{T^*}$. Thus, at the temperatures
employed in the simulations of Xu et al. ($T^*<10^{-3}$), the
combination $[T^*/p^*\lambda^3_\nu(T^*)]$ is essentially the
temperature-to-pressure ratio $[T^*/p^*]$, employed in Ref.
\cite{xu}.

In summary, we have illustrated the accuracy of the density-temperature-softness scaling of the dynamics of soft-sphere liquids, and shown that it provides a simple and useful conceptual tool to understand, within a unified framework, the phenomenology of both, thermally-driven molecular glass formers and density-driven hard-sphere--like colloidal liquids.  There are, of course, important pending assignments, such as the characterization of the non-equilibrium dynamics at, and beyond, the glass transition in this (HS) universality class, or the identification of other dynamic universality classes, particularly those influenced by the presence of attractive interactions. We expect that the results discussed in this work, together with the non-equilibrium extension of the SCGLE theory \cite{nonEqSCGLE}, will facilitate the progress in these directions.

ACKNOWLEDGMENTS:  This work was supported by
the Consejo Nacional de Ciencia y Tecnolog\'{\i}a (CONACYT,
M\'{e}xico), through grants No. 84076 and
FMSLP-2008-C02-107543.

\end{document}